\documentclass[amsmath,amssymb,superscriptaddress]{revtex4-2}

\usepackage{bm}
\usepackage{graphicx}
\usepackage{dcolumn}
\usepackage{amsmath}
\usepackage[latin1]{inputenc}
\usepackage{graphicx, psfrag}
\usepackage{amssymb}
\usepackage[colorlinks=true, citecolor=blue, urlcolor = blue, linkcolor= red, bookmarks=true]{hyperref}
\usepackage{float}
\usepackage{amsmath}
\usepackage{amsfonts}
\usepackage{dcolumn}
\usepackage{hyperref}
\usepackage{subfigure}
\usepackage{pgfplots}
\usepackage{epstopdf}
\usepackage{booktabs}

\begin{document}

\title{Rotating traversable wormhole geometries in the presence of three-form fields}

\author{Takol Tangphati} 
\email[]{takoltang@gmail.com}
\affiliation{School of Science, Walailak University, Thasala, \\Nakhon Si Thammarat, 80160, Thailand}

\author{Butsayapat Chaihao}
\email[]{butsayapat5@gmail.com}
\affiliation{Department of Mathematics and Computer Science, Faculty of Science, Chulalongkorn University,
Bangkok 10330, Thailand}

\author{Daris Samart}
\email[]{darisa@kku.ac.th}
\affiliation{Khon Kaen Particle Physics and Cosmology Theory Group (KKPaCT), Department of Physics, Faculty of Science, Khon Kaen University, Khon Kaen, 40002, Thailand}

\author{Phongpichit Channuie} 
\email[]{phongpichit.ch@mail.wu.ac.th}
\affiliation{School of Science, Walailak University, Thasala, \\Nakhon Si Thammarat, 80160, Thailand}
\affiliation{College of Graduate Studies, Walailak University, Thasala, Nakhon Si Thammarat, 80160, Thailand}

\author{Davood Momeni} 
\email{dmomeni@nvcc.edu}
\affiliation{Northern Virginia Community College, 8333 Little River Turnpike, Annandale, VA 22003}

\date{\today}

\begin{abstract}
In this work, we study the rotating wormhole geometries supported by a three-form field. We demonstrate for particular choices of parameters that it is possible for the matter fields threading the wormhole to satisfy the null and weak energy conditions throughout the spacetime, when the three-form field is present. In this case, the form field is interpreted as supporting the wormhole and all the exoticity is confined to it. Thus, the three-form curvature terms, which may be interpreted as a gravitational fluid, sustain these wormhole geometries. Additionally, we also address the ergoregion of the solutions.
    
\end{abstract}

\maketitle

\section{Introduction}

General Relativity (GR) permits the existence of traversable Lorentzian wormholes, which were first proposed by Ellis \cite{Ellis:1973yv,Ellis:1979bh} and Bronnikov \cite{Bronnikov:1973fh}. Basically, traversable Lorentzian wormholes necessitate the existence of an exotic matter field, which is coupled to gravity. This field features a kinetic term with the reverse sign, resulting in an energy-momentum tensor that violates the null energy condition \cite{Morris:1988cz,Morris:1988tu}. In recent times, researchers have explored the inclusion of phantom fields in cosmology due to their potential to drive accelerated expansion of the Universe \cite{Lobo:2005us}. However, alternative scenarios involving gravity theories with higher curvature terms have been considered, which allow for the construction of wormholes without the need for the exotic field, e.g., \cite{Lobo:2005us,Ghoroku:1992tz,Fukutaka:1989zb}.

Several types of exotic matter have been investigated in the context of traversable wormholes. One approach involves utilizing modified theories of gravity to create effective exotic fluids that can support the wormhole's throat. More specifically, Casimir energy has been considered as a potential source to generate a Traversable Wormhole \cite{Garattini:2019ivd}. It is used to proof the existence of negative energy which can be built in the laboratory. Its extension has been recently investigated by many authors, e.g., \cite{Jusufi:2020rpw,Garattini:2021kca,Samart:2021tvl,Garattini:2023qyo}.

From an observational astrophysical perspective, efforts have been made to search for wormholes \cite{Abe:2010ap,Toki:2011zu,Takahashi:2013jqa}. These enigmatic structures have been investigated as potential gravitational lenses \cite{Cramer:1994qj,Perlick:2003vg}, with particular attention given to studying their Einstein rings \cite{Tsukamoto:2012xs}, and shadows \cite{Bambi:2013nla,Nedkova:2013msa}. Most studies thus far have focused on static wormholes, although astrophysical objects typically exhibit rotation. Hence, understanding the characteristics of rotating wormholes is of great interest. Additionally, it is worth noting that the static Ellis wormholes in General Relativity (GR) are known to be unstable \cite{Shinkai:2002gv,Gonzalez:2008wd,Gonzalez:2008xk,Torii:2013xba}, and the introduction of rotation might offer a possibility of stabilizing them \cite{Matos:2005uh,Dzhunushaliev:2013jja}. Although many aspects of these rotating wormholes have already been examined, one crucial characteristic that remains to be investigated is their stability.

In this work, we consider the three-form field to be responsible for supporting the rotating wormhole geometries. It was demonstrated that all the exoticity is confined to it \cite{Barros:2018lca}. Moreover, three-form fields \cite{Wongjun:2017spo,Koivisto:2009ew} are widely used in the literature and seem to present viable solutions to cosmological scenarios, e.g., \cite{Morais:2016bev,Koivisto:2012xm,Koivisto:2009fb,Koivisto:2009sd,SravanKumar:2016biw,DeFelice:2012jt,DeFelice:2012wy,Kumar:2014oka}. As mentioned in Ref.\cite{Barros:2018lca}, the three-form curvature terms, which may be interpreted as a gravitational fluid, sustain these wormhole geometries. Here the authors are essentially interested in finding wormhole geometries supported by three-forms, where the matter threading the wormhole satisfies the energy conditions. Noticed that very recent study on the deflection angle of light by traversable wormholes, which are supported by the three-form fields was carried out in Ref.\cite{Samart:2022atm}.

The plan of the work is structured as follows: In Sec. \ref{sec2}, we review the rotating traversable wormhole and its proper metric tensor form. We then introduce the three-form field to sustain rotating wormhole geometries. The properties of the wormhole; the flaring-out condition and the asymptotic flatness are investigated. Additionally, the gravitational and three-form field equations for the rotating traversable wormhole are presented. In Sec. \ref{sec3}, we present the arbitrary functions for the wormhole construction. The null and weak energy conditions are studied. Since the traversable wormhole is rotating around the axial axis, the ergoregion has occurred at some point of rotation speed. We also discuss about this feature. In Sec. \ref{sec4}, we conclude our findings.

\section{Rotating metric tensor \& wormholes supported by three-form field} \label{sec2}

In this section, we consider the spacetime describing a rotating object. As mentioned in \cite{Teo:1998dp}, the properties of the metric tensor are stationary and axially symmetric. This also means the spacetime has a time-like Killing vector field $\zeta^a \equiv \left( \partial / \partial t \right)^a$ which is invariant in time translation and has a space-like Killing vector field $\psi^a \equiv \left( \partial / \partial \varphi \right)^a$ which is invariant in rotating in an azimuths axis. According to Refs.\cite{Papapetrou,Carter1,Carter2}, the most general stationary and axisymmetric metric takes the form
\begin{eqnarray}
    ds^2 = g_{tt} dt^2 + 2 g_{t \psi} dt d \psi + g_{\psi \psi} d \psi^2 + g_{ij} dx^i dx^j,
    \label{metric_general}
\end{eqnarray}
where $i,j$ are the indices for the rest of the space-like coordinates. A time-dependent conformal factor in the Morris-Thorne wormhole metric tensor might be applied to prevent the energy conditions' violation; however, any observer travelling through this wormhole would experience the radius of wormhole increasing all directions which could not be practical traversable wormhole \cite{Teo:1998dp,Roman:1993,Kar:1994,Kar:1996}. In this work, we first consider the wormhole metric which describes a rotating wormhole spacetime in the spherical polar co-ordinates given by \cite{Teo:1998dp}
\begin{eqnarray}
    ds^2 &=& - e^{2\Phi(r)} dt^2 + \frac{dr^2}{1 - \frac{b(r)}{r}} + r^2 K(r)^2 \left[ d\theta^2 + \sin^2 \theta \left( d \varphi - \omega(r) dt \right)^2 \right], \label{metric_general2}
\end{eqnarray}
where $\Phi(r)$ is referred to as the redshift function which is associated with gravitational redshift. It is assumed to have a finite value across all points to prevent the formation of event horizons. This condition allows the wormhole to be traversable, according to references \cite{Butcher:2015sea}. Here $b(r)$ is denoted the shape function, as it depicts the form of the wormhole. The radial coordinate $r$ runs from a minimum value $r_0$, corresponding to the throat of the wormhole, where $b(r_0) = r_0$ at $r = r_0$. A key ingredient of wormholes is the so-called flaring-out condition \cite{Morris:1988cz}, given by $b(r) - b'(r) r \geq 0$, at the vicinity of the throat, where a prime denotes a derivative with respect to the radial coordinate $r$. Additionally, $b(r)/r \rightarrow 0$ as $r \rightarrow \infty$. Note that the additional condition $b(r)/r < 1$ is also imposed. $K(r)$ is a positive and non-decreasing function of $r$. It is worth mentioning that the above metric was first used by Hartle \cite{Hartle:1967ha,Hartle:1967he} in the study of relativistic rotating stars.

The asymptotic flatness is still required for the metric tensor at $r \rightarrow \infty$ where
\begin{eqnarray}
    \Phi(r) \rightarrow 0, \quad K(r) \rightarrow 1, \quad \omega(r) \rightarrow 0. \label{asymptotic_condition}
\end{eqnarray}
We choose the form of $\omega(r)$ to follow the asymptotic flatness \cite{Teo:1998dp}:
\begin{eqnarray}
    \omega(r) = \frac{2a}{r^3} + \mathcal{O} \left( \frac{1}{r^4}\right)\,,
\end{eqnarray}
where $a$ is the total angular momentum. The action of the 3-form field model to construct the wormhole reads \cite{Barros:2018lca}
\begin{eqnarray}
    \mathcal{S} &=& \int d^4 x  \sqrt{-g} \left( \frac{R}{2 \kappa^2} + \mathcal{L}_A \right) + \mathcal{S}_m, \label{action1}
\end{eqnarray}
where $g$ is the determinant of the metric tensor, $\kappa^2 \equiv 8 \pi G$, $R$ is the scalar curvature, $\mathcal{S}_m$ is the action of the ordinary mass and $\mathcal{L}_A$ is the Lagrangian density of the 3-form field described by
\begin{eqnarray}
    \mathcal{L}_A = - \frac{1}{48} F^2 + V(A^2),
\end{eqnarray}
where $F^2 = F^{\mu \nu} F_{\mu \nu}$ is the contraction of all indices of the 4-form strength tensor ($F = dA$)
\begin{eqnarray}
    F_{\alpha \beta \gamma \delta} = \nabla_{\alpha} A_{\beta \gamma \delta} - \nabla_{\beta} A_{\gamma \delta \alpha} + \nabla_{\gamma} A_{\delta \alpha \beta} - \nabla_{\delta} A_{\alpha \beta \gamma}
\end{eqnarray}

Varying the action in Eq.~(\ref{action1}) with respect to $A_{\alpha \beta \gamma}$, we obtain the field equation as
\begin{eqnarray}
    \nabla_{\alpha} F^{\alpha \beta \gamma \delta} = 12 \frac{\partial V}{\partial A^2} A^{\beta \gamma \delta} \,.\label{F_field_equation}
\end{eqnarray}
Practically, we are able to write the 3-form field $A_{\alpha \beta \gamma}$ in term of the 1-form field (vector) $B^{\delta}$ via
\begin{eqnarray}
    B^{\delta}  = \frac{1}{3!} \frac{1}{\sqrt{-g}} \epsilon^{\delta \alpha \beta \gamma} A_{\alpha \beta \gamma}, \label{B_A}
\end{eqnarray}
where we have considered a 4-dimensional spacetime and 3-form field, and setting $n = 4$ and $p = 3$ for this work. We can invert Eq.~(\ref{B_A}) to write the 3-form field in terms of its dual as shown
\begin{eqnarray}
    A_{\alpha \beta \gamma} = \sqrt{-g} \epsilon_{\alpha \beta \gamma \delta} B^{\delta}. \label{A_B}
\end{eqnarray}
We have a choice to choose the components of the vector $B^{\delta}$ \cite{Koivisto:2009sd}
\begin{eqnarray}
    B^{\delta} = \frac{\zeta(r)}{\sqrt{2}} \left(0, \left( 1 - \frac{b(r)}{r} \right)^{1/2} , 0, \frac{1}{r \sin \theta} \right)^{T}, \label{B_components}
\end{eqnarray}
where $\zeta(r)$ is an auxiliary function of the 3-form field in the metric tensor Eq.~(\ref{metric_general2}). We express the non-trivial components of the 3-form field 
\begin{eqnarray}
    A_{t \theta \phi} = A_{\phi t \theta} = A_{\theta \phi t} = - A_{t \phi \theta} = - A_{\theta t \phi} = - A_{\phi \theta t} = e^{\Phi(r)} r^2 \sin\theta\, \zeta(r). \label{A_components}
\end{eqnarray}
The above relations allow us to express $A^{2}$ of the 3-form fields as
\begin{eqnarray}
    A^2 = A_{\alpha \beta \gamma} A^{\alpha \beta \gamma} = -6 \zeta^2 (r). \label{AA_contraction}
\end{eqnarray}
It is noteworthy that, regardless of the angular component in the dual vector $B^{\delta}$ in Eq.~(\ref{B_components}), there is no effect of the angular part from metric tensor on the square of the 3-form fields in Eq.~(\ref{AA_contraction}). Now we consider the kinetic term of the Lagrangian density of the 3-form field $\mathbf{K}(r)$
\begin{eqnarray}
    \mathbf{K}(r) \equiv -\frac{1}{48} F^2 = -\frac{1}{48} F^{\alpha \beta \gamma \delta} F_{\alpha \beta \gamma \delta} = \frac{1}{2} \left( 1 - \frac{b(r)}{r} \right) \left[ \zeta(r) \left( \Phi'(r) + \frac{2}{r} \right) + \zeta'(r) \right]^2.
\end{eqnarray}
Owing to the fact that the angular part does not appear in the square of the 3-form field, the kinetic term of the 3-form field still has no angular part at all (see the Ref. \cite{DeFelice:2012jt}). Also note that the kinetic term will diminish at the throat of the wormhole $r = r_0 = b(r_0)$. Now we vary the action in Eq.~(\ref{action1}) with respect to the metric tensor $g^{\mu \nu}$ and obtain the field equations
\begin{eqnarray}
    G_{\mu \nu} &=& 8 \pi T_{\mu \nu}^{(\text{eff})} = 8 \pi \left( T_{\mu \nu}^{(\text{A})} + T_{\mu \nu}^{(\text{m})} \right),
\end{eqnarray}
where $T_{\mu \nu}^{(\text{A})}$ is the energy momentum tensor of the 3-form field, $T_{\mu \nu}^{(\text{m})}$ is the energy momentum tensor of matter. The energy momentum tensor of the 3-form field can be expressed to obtain
\begin{eqnarray}
    T^{(\text{A}) \mu} {}_{\nu} = \frac{1}{6} F^{\mu \alpha \beta \gamma} F_{\nu \alpha \beta \nu} + 6 \frac{\partial V}{\partial A^2} A^{\mu \alpha \beta} A_{\nu \alpha \beta} + \mathcal{L}_A \delta^{\mu} {}_{\nu}.
\end{eqnarray}
The energy momentum tensor of 3-form field in the rotating wormhole metric has non-trivial components as follows
\begin{eqnarray}
    T^{(\text{A}) t} {}_{t} &=& -\rho_A = -V + \frac{\partial V}{\partial \zeta} \zeta - \mathbf{K}\,, \\
    T^{(\text{A}) r} {}_{r} &=& p_{r,A} = -V + \mathbf{K} \,,\\
    T^{(\text{A}) \theta} {}_{\theta} &=& p_{\theta, A} = -V + \frac{\partial V}{\partial \zeta} \zeta - \mathbf{K} \,,\\
    T^{(\text{A}) \phi} {}_{\phi} &=& p_{\theta, A} = -V + \frac{\partial V}{\partial \zeta} \zeta - \mathbf{K}\,.
\end{eqnarray}

The gravitational field equations
\begin{eqnarray}
    \rho_{\text{eff}} &=& \rho_m + \rho_A \nonumber \\
    &=& \frac{e^{-2 \Phi}}{4r^2} \left[ -b' \left( 4 e^{2 \phi} + r^3 \sin^2 \theta \, \omega \omega' \right) + r^2 \sin^2 \theta \left( b \omega \omega' +  \left( r - b \right) \left( r \omega'^2 + \omega \left(8 - 2r \Phi' \right) \omega' + 2 r \omega'' \right) \right) \right], \label{rh}\\
    p_{r, \text{eff}} &=& p_{r, \text{m}} + p_{r, \text{A}} \nonumber \\
    &=& -\frac{b}{r^3} + \frac{1}{4r} \left( 1 - \frac{b}{r}\right) \left( 8 \Phi' + e^{-2 \Phi} r^3 \sin^2 \theta \omega'^2 \right),\label{pr} \\
    p_{\theta, \text{eff}} &=& p_{\theta, \text{m}} + p_{\theta, \text{A}} \nonumber \\
    &=& \left(1 - \frac{b}{r} \right) \left( \frac{\Phi'}{r} + \Phi'^2 + \Phi'' - \frac{e^{-2 \Phi}}{4} r^2 \sin^2 \theta \omega'^2 + \left( \frac{b - b'r}{2r^2 (r-b)} \right) + \left( \frac{b - b'r}{2r (r-b)} \right) \Phi' \right),\label{the} \\
    p_{\phi, \text{eff}} &=& p_{\phi, \text{m}} + p_{\phi, \text{A}} \nonumber \\
    &=& \left(1 - \frac{b}{r} \right) \left( \frac{\Phi'}{r} + \Phi'^2 + \Phi'' + \left( \frac{b - b'r}{2r^2 (r-b)} \right) + \left( \frac{b - b'r}{2r (r-b)} \right) \Phi' \right) + \Delta,\label{thi}
\end{eqnarray}
where
\begin{eqnarray}
    \Delta &=& \frac{1}{4} \left( e^{-2 \Phi} \sin^2 \theta \left[ \omega' \left( \omega \left(7b + r (b'-8) + 2r (r - b) \Phi' \right) + 3 (b-r)r \omega' \right) + 2 (b-r) r \omega \omega'' \right] \right)\,.\nonumber
\end{eqnarray}

The field equation of the three-form field in Eq.~(\ref{F_field_equation}) in the rotating wormhole metric tensor reads
\begin{eqnarray}
    && 2r^2 \frac{\partial V}{\partial \zeta} + \zeta' \left( \frac{4r^2 - 3r b - r^2 b'}{r} + 2r (r-b) \Phi' \right) + 2r (r - b) \zeta'' \nonumber\\&+& \frac{\zeta}{r} \left( -4r + 6 b - 2r b' + r \Phi' ( b - r b) 2r^2 (r-b) \Phi'' \right) = 0.\label{zet}
\end{eqnarray}
The above relation imposes an additional constraint on the unknown functions and is significantly useful in solving explicit wormhole solutions.

\section{Energy conditions and Ergoregions} \label{sec3}
In order to find wormhole solutions, we will specify the redshift and shape functions, and assume further a form for $\zeta$. In this work, we follow the work done by Ref.\cite{Barros:2018lca}. Additionally, in Refs \cite{Capozziello:2013vna,Capozziello:2014bqa,Capozziello:2018wul}, the energy momentum tensor of ordinary matter holds the energy conditions where as the 3-form field involves the violation of NEC and WEC. We need to solve the five independent equations, which consist of three gravitational field equations Eqs.(\ref{rh})-(\ref{thi}) and the equation of motion for $\zeta$, i.e., Eq.(\ref{zet}). Following the notation of \cite{Capozziello:2012hr}, we consider the metric functions of the form
\begin{eqnarray}
    b(r) &=& r_0 \left( \frac{r_0}{r} \right)^{\beta}, \\
    \Phi(r) &=& \Phi_0 \left( \frac{r_0}{r} \right)^{\alpha},
\end{eqnarray}
where $\beta > -1$, $\alpha > 0$, and for the $\zeta$ function with $\gamma > 0$:
\begin{eqnarray}
    \zeta(r) = \zeta_0 \left( \frac{r_0}{r} \right)^{\gamma}\,.\label{ze}
\end{eqnarray}
Note that Eq.(\ref{ze}) takes the value $\zeta=\zeta_0$ at the throat and tends to zero at spatial infinity. The analytic solution for $V$ takes the form
\begin{eqnarray}
    V(r) = \frac{\gamma \zeta_0^2}{2r^3} \left( \left( r_0 \left( \frac{r_0}{r} \right)^{\beta} - r \right)(\gamma - 2) + \left( \frac{r_0}{r} \right)^{\alpha} \Phi_0 \left( - \frac{2r \alpha (1 + \alpha + \gamma)}{2 + \alpha + 2 \gamma} \right) + \frac{r_0 \left( \frac{r_0}{r} \right) \alpha  (3 + 2 \alpha + \beta + 2 \gamma)}{3 + \alpha + \beta + 2 \gamma} \right) + c_1,
\end{eqnarray}
where $c_1$ in the integrating constant. Even though the model of traversable wormholes is one of various solutions of Einstein's general relativity, it suffers the violation of the energy conditions, i.e., null and weak energy conditions \cite{Samart:2021tvl, Banerjee:2021mqk}. Then the matter that can distort the spacetime to construct traversable wormholes is called exotic matter. In this work, we focus to synthesize the rotating traversable wormhole with the 3-form field without invoking the exotic matter . The null energy condition (NEC) states that the relation $T_{\mu \nu} k^{\mu} k^{\nu} \geq 0$ for all null vector field $\vec{k}$. The weak energy condition (WEC) is defined based on the measurement of the matter density from an observer which cannot be negative $T_{\mu \nu} U^{\mu} U^{\nu} \geq 0$ where $U^{\mu}$ is any time-like vector.

\begin{figure}[h]
    \centering
    \includegraphics[width = 8 cm]{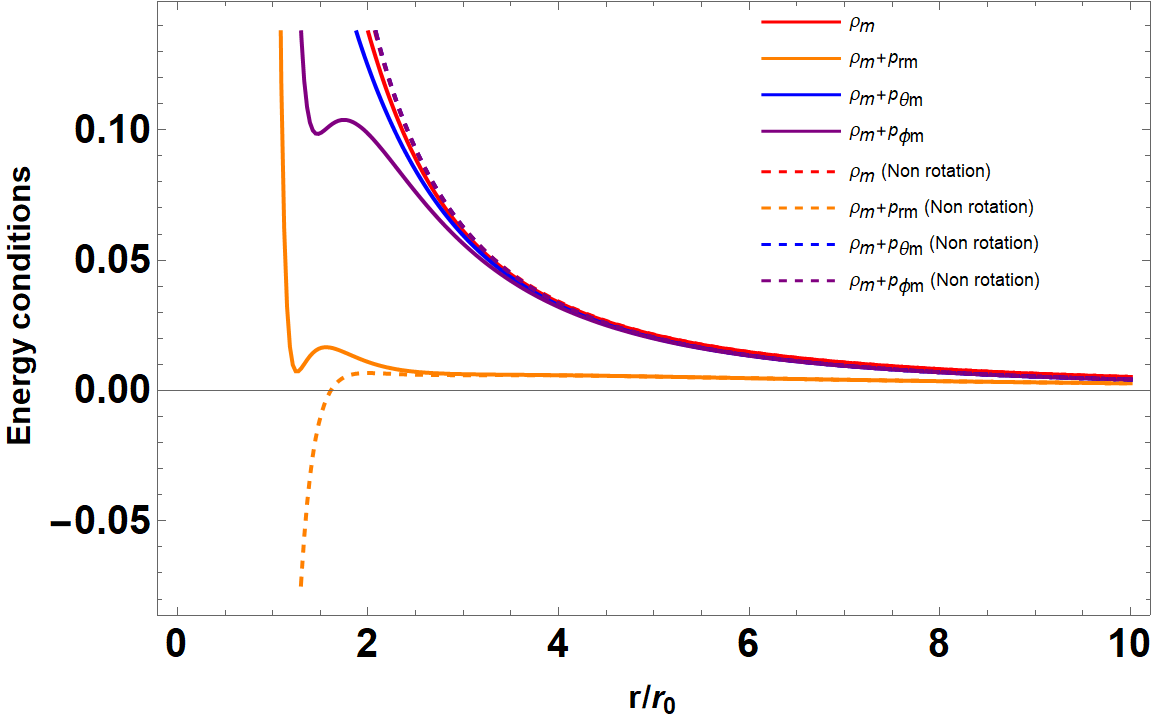}
    \includegraphics[width = 8 cm]{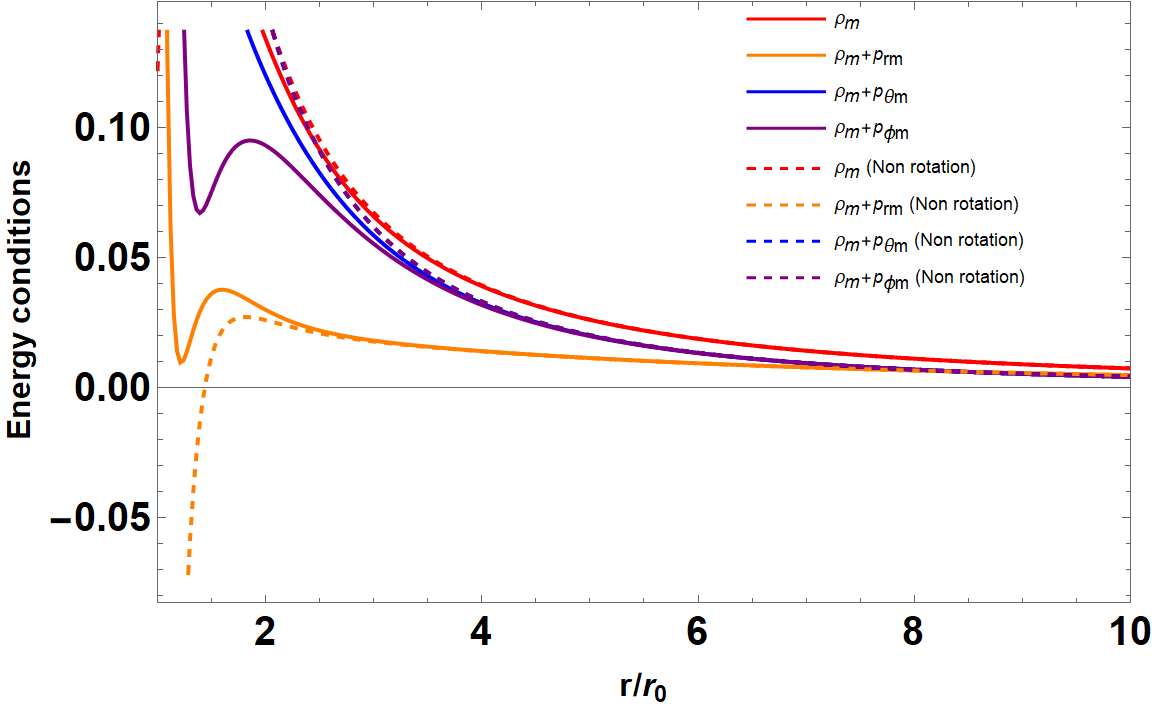}
    \includegraphics[width = 8 cm]{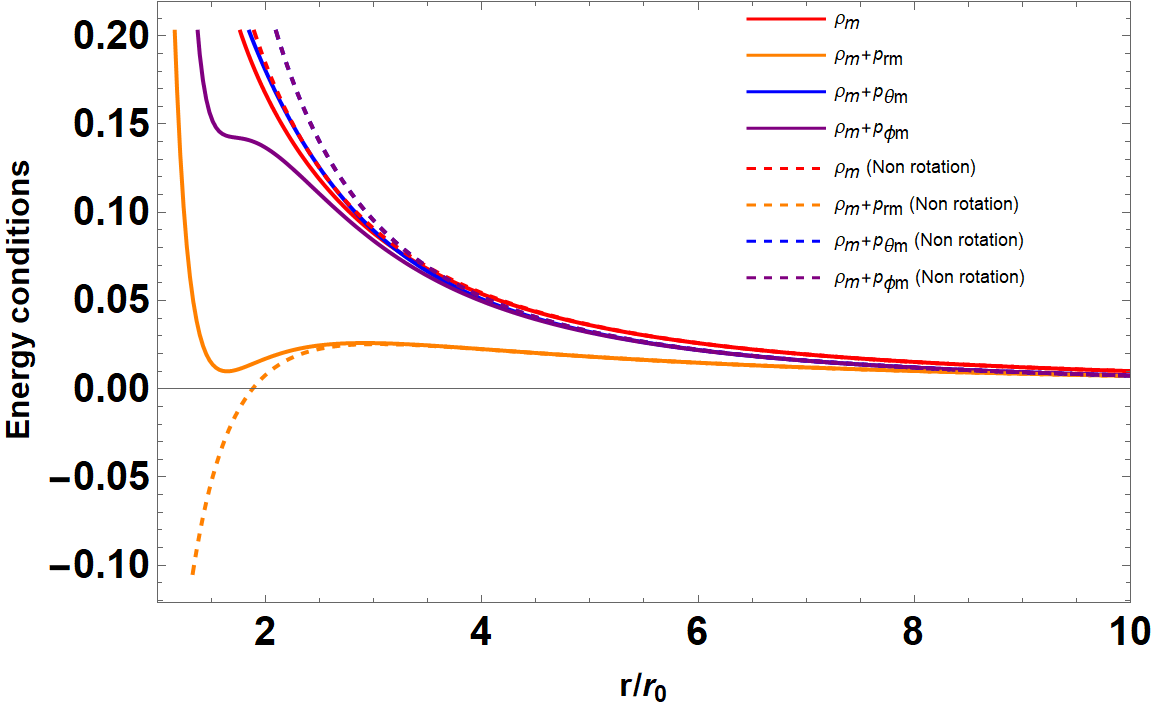}
    \includegraphics[width = 8 cm]{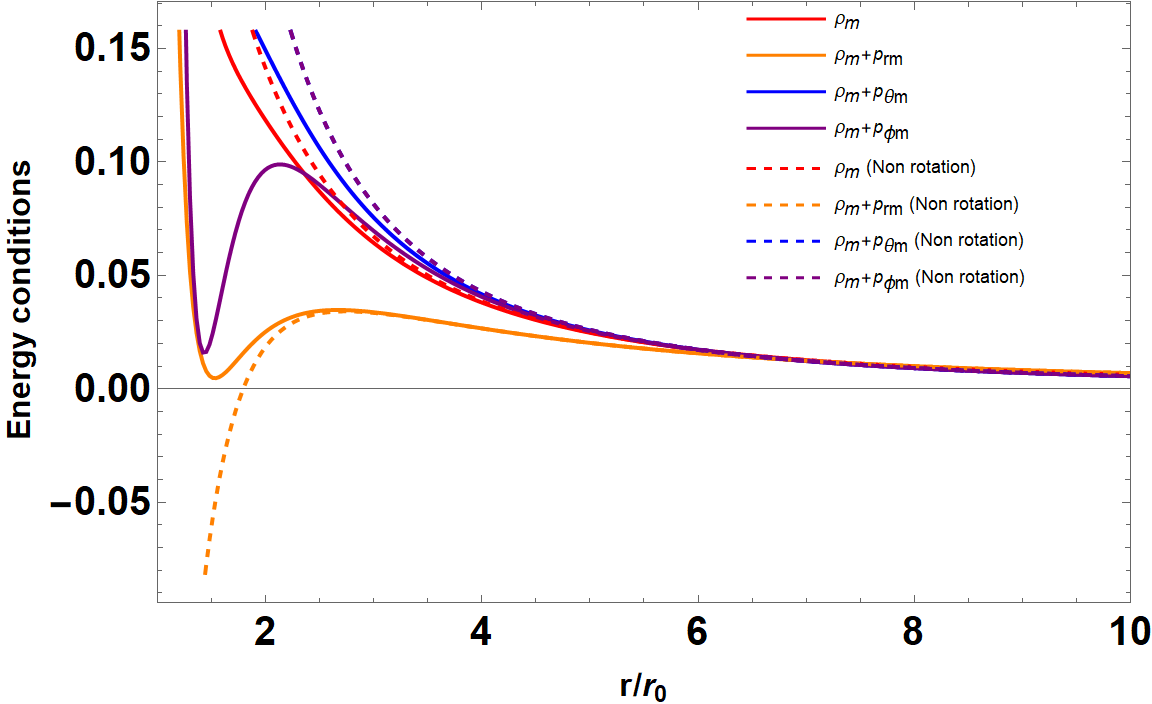}
    \caption{We display null and weak energy conditions of material for constructing the rotating wormholes supported by 3-form field with the parameter set: $\Phi = -0.6, \zeta_0 = 1.0, \alpha = 3.0, \beta = -0.7, \gamma = 1.0, c_1 = 0, a = 0.7, \theta = \pi/2$ (Upper-left panel), $\Phi = -0.7, \zeta_0 = 1.0, \alpha = 3.0, \beta = -0.7, \gamma = 0.5, c_1 = 0, a = 0.7, \theta = \pi/2$ (Upper-right panel), $\Phi = -1.0, \zeta_0 = 2.0, \alpha = 0.2, \beta = -0.9, \gamma = 1.0, c_1 = 0, a = 0.7, \theta = \pi/2$ (Lower-left panel) and  $\Phi = -0.95, \zeta_0 = 1.8, \alpha = 0.5, \beta = -0.8, \gamma = 1.5, c_1 = 0, a = 0.65, \theta = \pi/2$ (Lower-right panel), where the dashed lines represent the non-rotating cases ($a = 0$).
    }
    \label{Energy}
\end{figure}

In Fig.(\ref{Energy}), we demonstrate that the energy densities of specific solutions in which the matter component satisfies with both the NEC and WEC. This indicates that the presence of a three-form field is essential for maintaining the wormhole, and and all the exoticity of the object is confined to the field itself and the matter sources thread the wormhole without violating the NEC and WEC.

\begin{figure}[h]
    \centering
    \includegraphics[width = 12 cm]{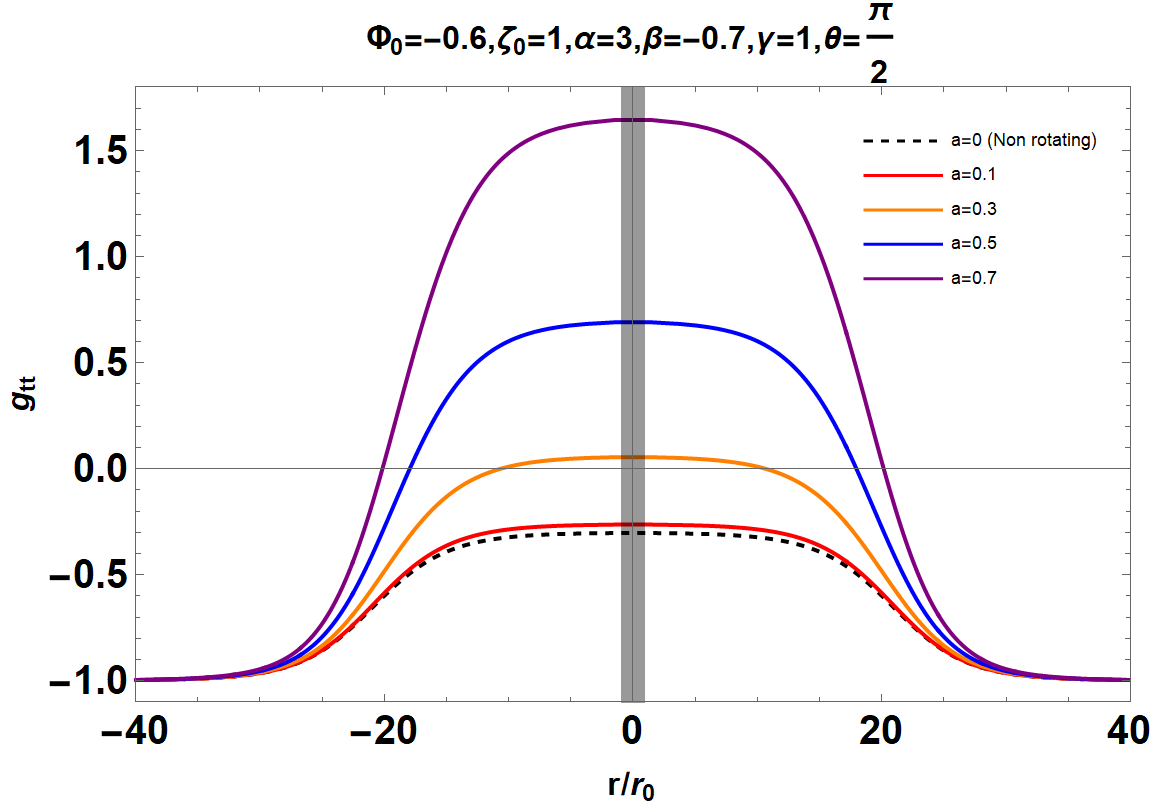}
    \caption{The components of $g_{tt}$ of the rotating wormhole with the 3-form field are presented with the variation of $a \in [0.0, 0.7]$. The diagram is split into two regions; $r/r_0 \geq 1$ (our universe) and $r/r_0 \leq -1$ (the other universe) where the non-exist region is between $-1 < r/r_0 < 1$. Note that all cases in the left panel satisfy NEC and WEC. These cause the ergoregion like the rotating black hole while the small rotating wormholes in the right panel do not cause the ergoregion ($a = 0$ and $a = 0.1$). However, WEC and NEC are not satisfied for such cases.}
    \label{fig:gtt}
\end{figure}

\begin{figure}[h]
    \centering
    \includegraphics[width = 8 cm]{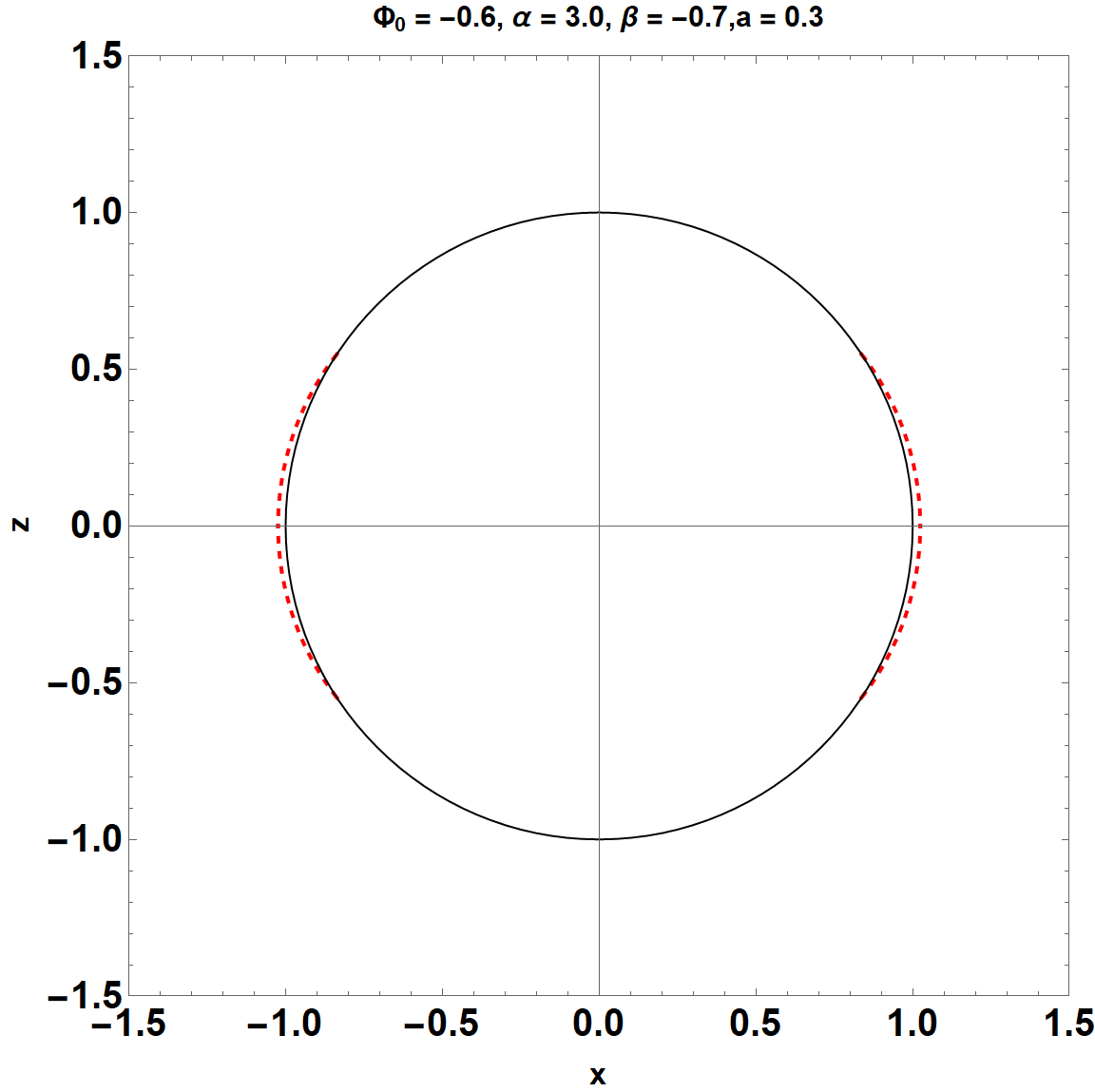}
    \includegraphics[width = 8 cm]{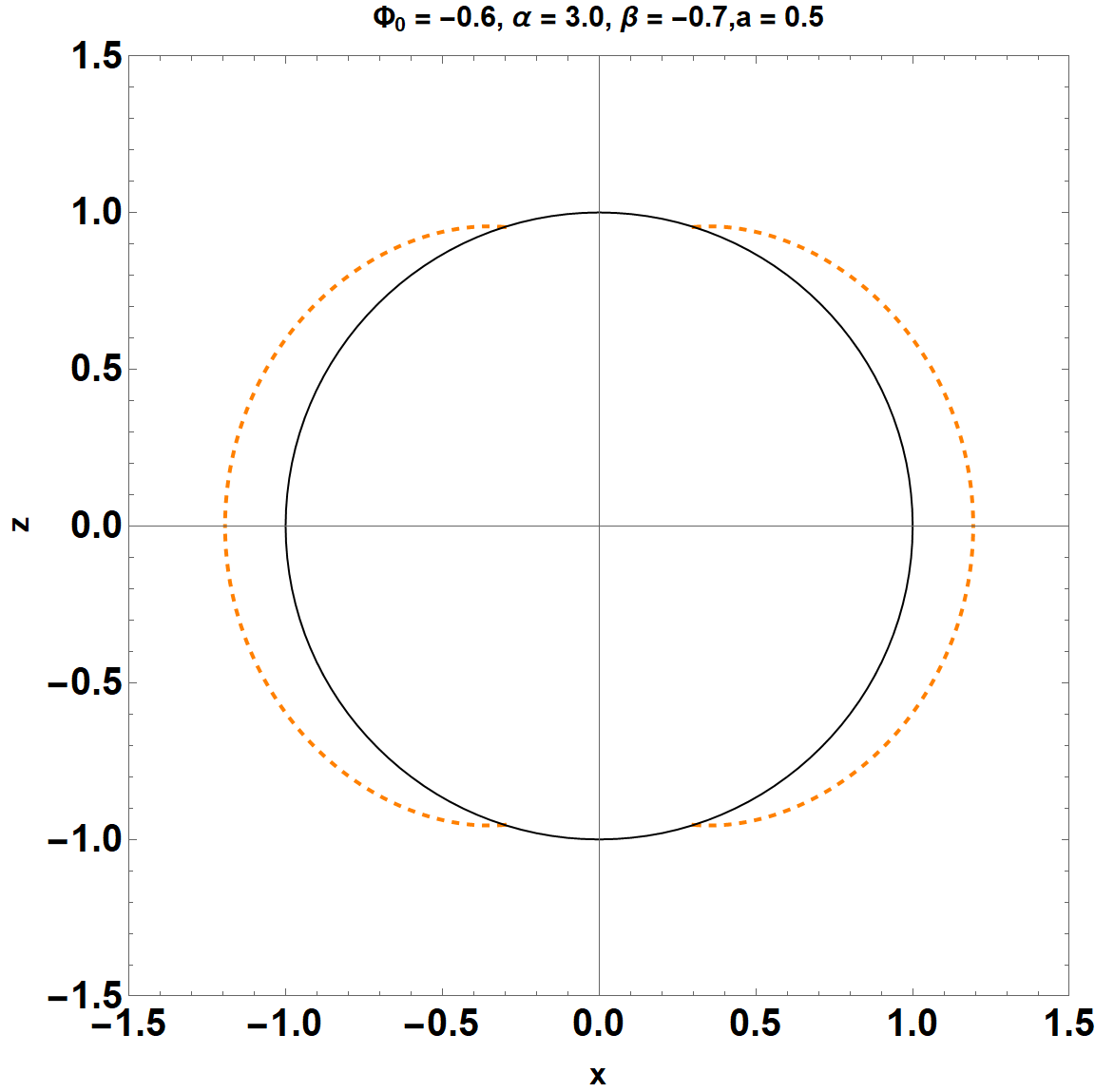}
    \includegraphics[width = 8 cm]{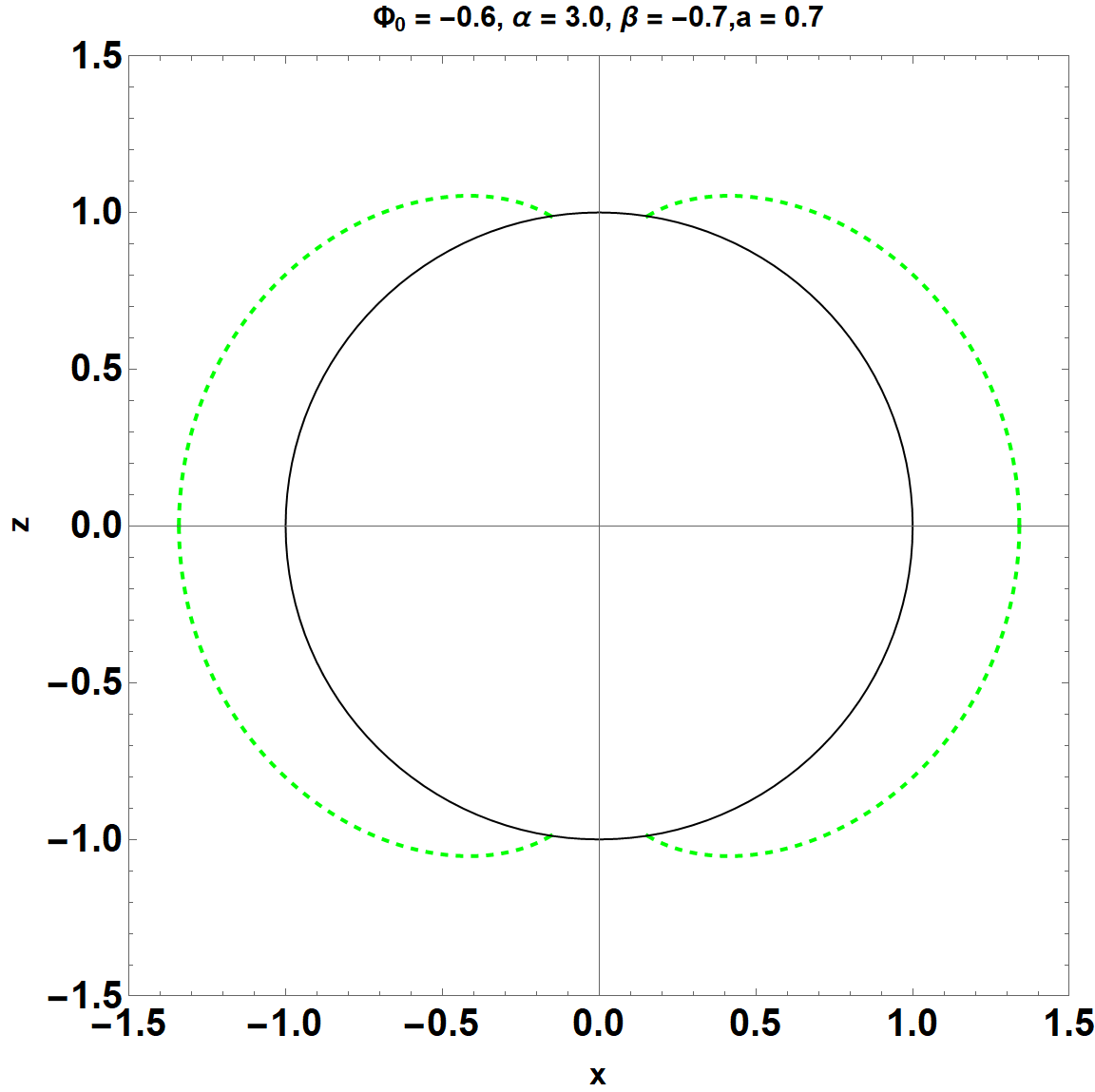}
    \includegraphics[width = 8 cm]{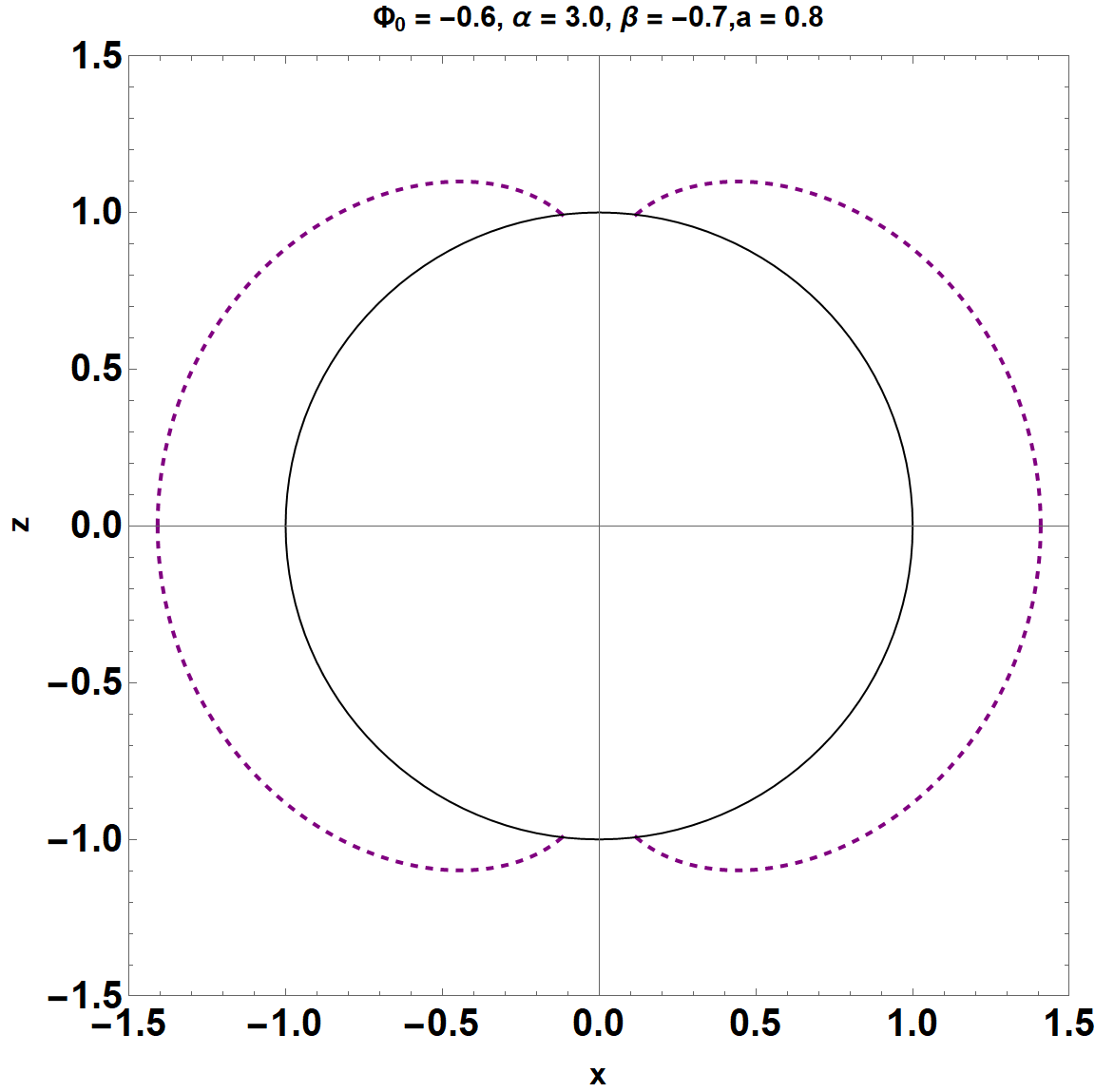}
    \caption{The ergoregions of the rotating wormhole with the three-form field are presented in the dashed color curves with the variation of $a \in [0.3,0.8]$ using the parameter set: $\Phi_0 = -0.5$, $\alpha = 3.0$, and $\beta = -0.7$. All of these cases satisfy NEC and WEC.}
    \label{fig:ergoregion}
\end{figure}

If the speed of the wormhole's rotation is high enough, $g_{tt}$ becomes positive in a certain area beyond the throat, suggesting the existence of an ergoregion where particles can no longer remain stationary concerning infinity. The ergoregion of the solutions is defined as the region where the time-time component of the metric is positive, $g_{tt} > 0$. Its boundary is referred to as the ergosurface where $g_{tt} = 0$. In our work, an ergoregion of a rotating wormhole can be determined when
\begin{eqnarray}
g_{tt} = - e^{2\Phi(r)} + r^2 K(r)^2 \omega(r)^2 \sin^2 \theta \geq 0\,, 
\end{eqnarray}
and the ergosurface by $g_{tt} = 0$ \cite{Shaikh:2018kfv,Patel:2022jbk}. The ergosurface for the metric is given by
\begin{eqnarray}
g_{tt} = - e^{2\Phi(r)} + r^2 K(r)^2 \omega(r)^2 \sin^2 \theta = 0\,, \label{gtt}
\end{eqnarray}
Since the ergoregion doesn't extent up to the poles $\theta=0$ and $\theta=\pi$, there exist a critical angle $\theta_{c}$, where the ergosphere exists in between $\theta_{c}$ and $\pi-\theta_{c}$, for all $0<\theta_{c} \leq \pi/2$. This critical angle can determined at the throat of the wormhole using Eq.(\ref{gtt}) as
\begin{eqnarray}
\sin \theta_{c}=\Big|\frac{e^{\Phi_{0}}}{r_{0}\, K_{0} \omega_{0}}\Big| \,. \label{gtt1}
\end{eqnarray}
Moreover, the presence of the ergosphere relies on the spin parameter surpassing a crucial threshold $a_{c}$, which corresponds to $\sin\theta_{c}=1$ or $\omega_{c} = 2a_{c}/r^{3}_{0}=e^{\Phi_{0}}/r_0 K_0$. When considering the wormhole metric (\ref{gtt}) with $r_0 = 1.0,\,K_{0}=1$ and $\Phi_{0}=-0.6$, the critical value is $a_{c} = 0.274406$, see also Lorentzian traversable wormholes \cite{Rahaman:2021web}. Fig.\ref{fig:gtt} illustrates the ergosphere's behavior in the equatorial plane by varying the angular momentum $a$. Here the ergoregion increases with increasing $a$. The ergosphere for these values is displayed in Fig.\ref{fig:ergoregion}.

Rotating objects, e.g., black holes, are known to be spending their rotational energy on amplification of incident waves of perturbation. This phenomenon occurs also for various rotating compact bodies, such as, for example, conducting cylinders, and is called superradiance \cite{Konoplya:2010kv,Franzin:2022iai,Mazza:2021rgq}. When considering rotating traversable wormholes, one could probably expect that the same superradiance should take place. However, it was shown that rotating axially symmetric traversable wormholes do not allow for the superradiance. The situation is similar to that of Teo's rotating wormhole example \cite{Teo:1998dp}. However, along the line of the present work, a phenomenology of superradiance emerging from rotating traversable wormholes is still underway.

\section{Conclusions} \label{sec4}
In this work, we have investigated the solutions of the rotating traversable wormhole interacting with the three-form field. The stationary and axisymmetric metric in the spherical polar co-ordinates has been adopted in this work. We have demonstrated that the asymptotic flatness and the flaring-out condition are satisfied. We obtained the field equation of the curved spacetime in the rotating traversable wormhole geometries which is the extension from the traditional static traversable wormholes \cite{Barros:2018lca}. We have considered the shape function and the red shift function for the traversable wormhole and an arbitrary function for three-form field proposed by Ref.\cite{Barros:2018lca}. This allows us to obtain the numerical solutions. Our results showed that the energy conditions such as NEC and WEC are satisfied. This is so since the three-form field behaves as a gravitational fluid to sustain the wormhole geometries. 

Furthermore, we have shown that using particular choices of parameters the existence of the ergoregion of the rotating traversable wormhole is possible. We have estimated the critical value of the angular momentum $a_{c}$ for which the ergoregion can emerge. We found that the ergoregion of a rotating wormhole increases with increasing $a$ implying that the emergence of the ergoregion of the wormhole strongly depends on the speed of the wormhole rotation. We have displayed the ergoregions of the rotating wormhole with the three-form field using the parameter set: $\Phi_0 = -0.5$, $\alpha = 3.0$, and $\beta = -0.7$ as an example. Note that all of these cases satisfy NEC and WEC. Along the line of the present work, the study of the deflection angle of light by this traversable wormholes supported by the three-form fields is possible. Additionally, the photon geodesic motion under the effective potential of the rotating wormhole background is worth investigated. The radius of the photon sphere is a very useful observable used to analyze the geometrical structures of a wormhole. It is widely known that the appearance of a shadow is a phenomenon which is not restricted only to black hole spacetimes. Therefore, the shadow of this class of rotating traversable wormholes is also an interesting phenomena, see e.g., \cite{Nedkova:2013msa}. We leave these interesting topics for our ongoing investigation.

\section*{Acknowledgments}

T. Tangphati is financially supported by Research and Innovation Institute of Excellence, Walailak University, Thailand under a contract No. WU66267. The work of P. Channuie is financially supported by Thailand NSRF via PMU-B under grant number PCB37G660013

\end{document}